\documentclass[twocolumn]{svjour3}     
\usepackage[english]{babel}
\usepackage[caption=false]{subfig}
\usepackage{amsmath,amssymb,graphicx,float}
\usepackage{hyperref}
\newcommand{\be}{\begin{equation}}
\newcommand{\ee}{\end{equation}}
\newcommand{\ea}{\end{eqnarray}}
\newcommand{\ba}{\begin{eqnarray}}

\newcommand{\wt}{\widetilde}
\newcommand{\xx}{\boldsymbol{x}}
\newcommand{\y}{\boldsymbol{y}}

\begin{document}

\title{An alternative Noncommutative approach to diagnose the Higgs boson degeneracy at 125~GeV}

\author{M. A. De Andrade \and C. Neves}
\institute{M. A. De Andrade \at
			Departamento de Matem\'{a}tica, F\'{\i}sica e Computa\c{c}\~{a}o, Faculdade de Tecnologia, \\ Universidade do Estado do Rio de Janeiro,\\
Rodovia Presidente Dutra, Km 298, P\'{o}lo Industrial, CEP 27537-000, Resende-RJ, Brazil.\\
\email{marco@fat.uerj.br}
\and
	C. Neves\at
	Departamento de Matem\'{a}tica, F\'{\i}sica e Computa\c{c}\~{a}o, Faculdade de Tecnologia, \\ Universidade do Estado do Rio de Janeiro,\\
Rodovia Presidente Dutra, Km 298, P\'{o}lo Industrial, CEP 27537-000, Resende-RJ, Brazil.\\
\email{clifford@fat.uerj.br}
}
\date{Received: date / Accepted: date}

\maketitle

\begin{abstract}
Nowadays, the CMS and ATLAS experimental results, for the Higgs boson mass, present the most advanced hardware, methodology and high precision statistical errors, then the Higgs mass is well established and consistent with Standard Model(SM). However, it is possible to argue that the CMS and ATLAS experiences are not yet enough to answer some issues present in the SM: some experimental results for the Higgs mass are off of statistical errors range and the hierarchy problem. Hence, there is some room to explain the Higgs mass degeneracy issue in an alternative way. In this work, this issue is investigated at an alternative noncommutative(NC) framework.
The role played by an alternative noncommutativity -- associated with damping phenomenon --, based on non-Moyal $\ast$-product method, in a $O(2)$ scalar field theory is investigated; precisely, how this noncommutativity affects
the spontaneous symmetry breaking (SSB) and Englert-Brout-Higgs-Guralnik-Hagen-Kibble(EBHGHK) mechanism. 
This subject is explored in a three-fold way: (a) it is shown that vacuum expectation value(VEV) and Higgs boson mass depend on the NC parameter, \textit{i.e.}, they are parametrized by the noncommutativity; (b) the gauge fields gain masses that, now, are also parametrized by the  noncommutativity; (c) we argue that the noncommutativity, which is a damping-like phenomenon, might be interpreted as being due to the Higgs couplings.
\end{abstract}


\section{Introduction}

\label{sec:intro}

Although the latest CMS and ATLAS experimental results for the Higgs boson mass are consistent with the SM and, respectively, are given by $(125.38\pm0.14)$~GeV\cite{cms-latest} and $(124.79\pm0.24)$~GeV\cite{atlas-higgs}, one cannot yet rule out the presence of a mass degenerate scalar state at 125 GeV \cite{gunion1,gunion2,ferreira,drozd,grossman1,munir,grossman2,king,david,moretti1,wang,carena,moretti2,bian,cabrera,haber}. Further, in the Higgs boson phenomenology, it seems that there is a Higgs boson mass degeneracy close at those CMS and ATLAS Higgs boson masses, which might be due to coupling of Higgs boson with others particles: different types of bosons(W\cite{atlasW,cms5} and Z\cite{atlas3,atlas4}), heavier fermions (tau leptons)\cite{cms-tau,atlas-cms}, heaviest quarks (top\cite{higgs-top} and bottom\cite{higgs-botton,atlasB}) and muons\cite{CMS2b,atlas-muon,CMS2c,CMS2a}. In addition, there are two remarkable questions, among others, that deserve to be answered within the SM: (a) the hierarchy problem, why the Higgs boson is much lighter than the characteristic scale of gravity? (b) the existence of dark matter (DM) is not accommodate by SM, indeed
there is no viable DM candidate in the SM. So already these facts
ask for an alternative physics approach to handle with these issues. Recently, some authors\cite{carmona} extended the SM with a warped
extra dimension together with a new $\mathbb{Z}_2$-odd scalar singlet. There, it was shown that the scalar excitations necessarily mix with the Higgs boson and, consequently, leads to modifications of the Higgs couplings. Moreover, mass-degenerate Higgs scenarios within the Next-to-Minimal Supersymmetric Standard Model (NMSSM) were addressed\cite{gunion1,gunion2,munir,moretti1} considering two quasi-degenerate Higgs states for the real and complex NMSSM, with a mass difference large enough to use the narrow width approximation, while Ref.\cite{moretti2} investigated the interference effects beyond the narrow width approximation. Besides, a statistical test based on the analysis of a signal strength matrix was proposed to demonstrate that results obtained from SM are due to the existence of more than one resonance\cite{grossman1,david}. In this context, a recent article\cite{cabrera} used the NMSSM doublet-singlet mixing structure\cite{king,carena,moretti2} of the Higgs sector to investigate the two mass-degenerate states hypothesis, which with different coupling structures combined simulate a single Higgs features. In another recent article\cite{haber}, the implications of a Higgs boson mass degeneracy in the scalar sector, among two (or more) scalar states of an extended Higgs sector, was investigated and it was shown the interplay between symmetries and natural mass degeneracies in the scalar sector.

Despite of the  highest CMS and ATLAS precisions, advanced hardware and methodology, it is still possible to suspect that the Higgs boson mass degeneracy is not detected at present CMS and/or ATLAS experiments because they demand a hardware and/or methodology improvement. So, there is some room for speculating a solution through alternative points of view, unlike to what was discussed before in those fuzzy mass-degenerate Higgs boson states scenario. With the aim to  tackle the mass-degenerate issue from another point of view, this article proposes to investigate the Higgs boson mass degeneracy in a toy model through an alternative noncommutative framework whose the induced noncommutativity is among the momenta, which avoids no-go problem\cite{CPJT}. Further, it is shown that the NC contribution in the model, which plays the role of a damping feature\cite{MACN1}, might work out as the Higgs couplings, which might explain the Higgs boson mass degeneracy. 

Noncommutativity has been extensively investigated in different contexts: quantization procedure\cite{wigner,HG,MOYAL,mezincescu,Chaichian2001,bopp2,djemai1,djemai2,ANO2,ANO2b,ANO2a,MACN}, the Yang-Mills theory on a NC torus\cite{conne,CDS}, matrix model of M-theory\cite{banks1,LS,banks3,banks2}, string theory\cite{jabbari,jabbari1,DH,jabbari2,ardalan,jabbari4,jabbari5,SW} and D-brane\cite{jabbari3,schomerus,kont,CF,BS,MACN1}, SSB and EBHGHK mecha-\\nisms\cite{campbell,Chaichian2003,girroti2003,castorina2003,castorina2005,everton}. At these last scenario, it was investigated how noncommutativity affects the IR-UV mixing and the appearance of massless excitations\cite{campbell,girroti2003}, the relation between symmetry breaking in NC cut-off field theories\cite{castorina2003,castorina2005} and the role played by the noncommutativity in the masses generation of new boson\cite{everton}. Despite of all this extensive research, the whole role played by noncommutativity in mass generation, even in a simplest $O(2)$ scalar field theory, was not yet properly investigated because it is principally based on the Moyal $\ast$-product, \textit{i.e.}, based on the space-time noncommutativity or on  phase space coordinates that present a no-go problem\cite{CPJT}. With the aim to fill some gap into this matter and avoid the no-go problem, the NC induced in the present work is among the momenta. This strategy also allowed us to relate noncommutativity\cite{MACN1} with damping phenomenon. At this point, we would like to relate damping phenomenon with deep inelastic scattering\cite{FT1,FT2}. Deep inelastic scattering asserts that a large part of the kinetic energy of the incident nucleus is converted into internal energy of the nuclear system(dynamical mass increases) and, therefore, the kinetic energy of reaction products is a small fraction of that initial value. 
Based on this assumption, we conjecture that the Higgs boson -- mediating particle of Higgs field -- plays the role of incident nucleus, while a particle, coupled to Higgs field, is the nuclear system. In this context, we propose to explain Higgs boson mass degeneracy through damping phenomenon.

It is well known that, from the universe birth and until before SSB, the Higgs field is present throughout all space(the vacuum of outer space is not empty but contain this field) and the Higgs boson is the mediating particle of the proposed Higgs field. Following the ideas, previously presented, we propose that the Higgs field is coupling with an ensemble of particles and it works out as a damping system. Then, the Higgs boson mass depends on each type of coupling, which is represented by a damping coefficient, namely, a NC $\theta$-parameter. In this scenario, we can conjecture that the universe, \textit{ab initio}, might behavior as a damping system(NC system). In the NC universe genesis, the damping contribution might be null but, after that and, maybe, for a period of time, this non null contribution appears due to the coupling of the Higgs field with each particle belonging to the particle ensemble. Hence, the investigation should departure from the original damped system. In order to get this NC system, a damping feature(NC term dependent) is induced into the commutative system through Noncommutative Mapping\cite{MACN,MACN1}. After that, it is shown how noncommutativity affects the conception for SSB\cite{nambu,goldstone,lasinio1,lasinio2,gsw,anderson} and EBHGHK mechanism\cite{higgs1,englert,kibble2,higgs2,kibble1}. Consequently, it is proposed an alternative, perhaps speculative, theoretical explanation for the Higgs boson mass degeneracy near the latest CMS and ATLAS experimental results\cite{cms-latest,atlas-higgs}, which also contributes to change the masses gained by the gauge vector fields. 

This work is organized as follows. In section~\ref{sec:1}, we propose, through Noncommutative Mapping\cite{MACN}\footnote{In this article, the NC $\theta$-parameter is not, necessarily, small -- infinitesimal order -- since the NC Mapping method does not impose any condition on $\theta$-parameter, indeed, this method induces the noncommutativity in an exact form. On the order hand, the noncommutativity induced by the Moyal $\ast$-product bases on the expansion in order of $\theta$, then this parameter must be considered too small, \textit{i.e.}, it is  infinitesimal.}, a non- Moyal NC version for the classical $O(2)$ scalar field theory, indeed we get a damped $O(2)$ scalar field theory\cite{MACN1}. In section \ref{sec:2}, we investigate, at classical level, how the contribution of a damping feature affects the idea of SSB mechanism --  but we use quantum-mechanical language -- and, also, it is discussed how Higgs boson mass, near 125~GeV, might arise due to damping property. Further, it is also shown how it can be related with the damping phenomenon. In section \ref{sec:3}, the noncommutative contribution into EBHGHK mechanism is investigated and its consequence in masses gained by gauge fields is discussed. At the end, some conclusions are presented.

\section{NC scalar field theory}
\label{sec:1}

In order to illustrate the contribution of noncommutativity in the context of field theory, we chose a simplest scalar field in four space-time dimensions, namely, a global $O(2)$ scalar theory, whose its dynamics is governed by
\be
\label{lp42}
{\cal{L}}=\frac 12 (\partial_{\mu}\phi_i)(\partial^\mu\phi_i)-\frac{\mu^2}{2}\phi_i\phi_i-\frac {\lambda}{4} (\phi_i\phi_i)^2 ,
\ee
where $\lambda$ is a positive number, $\mu^2$ can be either positive or negative and the field $\phi_i$ transforms as a $2$-vector. The corresponding Hamiltonian is
\be
\label{hamiltoniana21}
{\cal{H}}=\frac{\pi_i\pi_i}{2}+\frac{(\nabla\phi_i)(\nabla\phi_i)}{2} + U,
\ee
where $\pi_i$ are the momenta conjugated to the fields $\phi_i$. The potential is given by
\be
\label{Cpotential}
U=\frac{\mu^2}{2}\phi_i\phi_i+\frac {\lambda}{4} (\phi_i\phi_i)^2 .
\ee
It is well know that if $\mu^2>0$, then the vacuum is at $\phi_i=0$ and the symmetry~$(\phi_i=-\phi_i)$ is manifest, and $\mu^2$ is the mass of the scalar modes. On the other hand, if $\mu^2<0$, then there is a new vacuum solution given by $(\phi_i\phi_i)={\frac{-{\mu}^2}{\lambda}}$, which has an infinite number of possible vacua. 

The main result on Ref.\cite{MACN} is that the commutative phase-space framework $(\phi_i,\pi_i)$, where the Poisson brackets relations are the canonical ones, can be mapped to the noncommutative phase-space framework $(\wt{\phi}_i, \wt{\pi}_i)$. In the latter, the noncommutativity is introduced into the present model deforming the Poisson brackets relations as
\ba
\label{ST045aaa}
\left\{\wt\phi_i(t,\xx)~,~ \wt\phi_j(t,\y)\right\} &=& 0,\nonumber\\
\left\{\wt\phi_i(t,\xx) ~,~ \wt\pi_j(t,\y)\right\} &=& \delta_{ij}\delta^{(3)}(\xx-\y),\,\, \\
\left\{\wt\pi_i(t,\xx) ~,~ \wt\pi_j(t,\y)\right\} &=& \theta \varepsilon_{ij}\delta^{(3)}(\xx-\y),\nonumber
\ea
where $\theta$ embraces the noncommutativity only among the momenta -- non-Moyal $\ast$- product --. Here, it is important to notice that is considered a noncommutativity that does not imply a no-go theorem, in opposition what was demonstrated in Ref.\cite{CPJT}.

At this point, we have to remember the 
 precise definition of locality given by Schwinger\cite{Schwinger}, 
\ba
\left[x^\mu,x^\nu\right]&=&0,\nonumber\\
\left[x^\mu, \psi^a(x)\right]&=&0,
\ea
so that
\be
\left\langle x|\psi^a(x)|x^\prime\right\rangle= \delta( x - x') \psi^a( x),
\ee
where $\psi^a(x)$ are a generic field operators functions of space-time coordinates. Despite of this, we should to quote D.V. Ahluwalia\cite{Ahluwalia}:
\begin{quote}
	 ``\ldots the fundamental assumption of locality in quantum field theory can only be considered as an approximation.'' 
\end{quote}
and, further, 
\begin{quote}
``\ldots essential result on non-commutativity of position measurements seems certain to survive when one or all assumptions of the setup considered are relaxed.''
\end{quote}
Since the noncommutativity proposed, Eq.(\ref{ST045aaa}), is not in position measure, nothing survives and, consequently, the fundamental assumption of locality survives, even when it is considered as an approximation.

In the commutative framework, the phase-space vector field is $\xi^\alpha=(\phi_i,\pi_i)$ and the associated symplectic matrix is
\be
f=\left(
\begin{array}{cc}
 0           ~&~   \delta_{ij} \cr
 -\delta_{ij}   ~&~  0 \cr
\end{array}
\right).
\ee
In the noncommutative framework, the phase-space vector field is $\wt\xi^\alpha=(\wt{\phi}_i, \wt{\pi}_i)$ and the associated deformed symplectic matrix is
\be
\label{ST050aaa}
{\wt{f}}=\left(
\begin{array}{cc}
 0 ~&~    \delta_{ij}  \cr
 -\delta_{ij}     ~&~  \theta \varepsilon_{ij}\cr
\end{array}
\right),
\ee
and those brackets given in Eq.(\ref{ST045aaa}) can be brought together through a single Poisson bracket,
\be
\{\wt\xi^\alpha(t,\xx)~,~\wt\xi^\beta(t,\y)\}=\wt{f}^{\alpha\beta}\delta^{(3)}(\xx-\y).
\ee
The NC transformation matrix\cite{MACN}, $\,R=\sqrt{\wt{f}\,f^{-1}}$, is written as
\be
\label{ST062aaa}
R=\left(
\begin{array}{cc}
 \delta_{ij}             ~&~   0  \cr
 \frac12\,\theta \varepsilon_{ij}    ~&~  \delta_{ij} \cr
\end{array}
\right).
\ee
As the commutative phase-space vector field $\,\xi^\alpha\,$ change to the NC one $\,\wt\xi^\alpha\,$ through $\,d\wt\xi^\alpha=R^{\alpha}_{~\beta}\,d\xi^\beta$, it follows that
\be
\label{q-trans1}
\wt{\phi}_i={\phi}_i~~,~~\wt{\pi}_i=\pi_i +\frac12\,\theta \varepsilon_{ij}\,{\phi}_j.
\ee
Here, it is important to stress that this noncommutativity does not follow the Moyal $\ast$-product, based on the Bopp shifts\cite{bopp,kubo}; note that the fields $(\phi_i)$ do not shift.

In agreement with the NC Mapping\cite{MACN}, the NC first-order Lagrangian can be read as
\be
\label{ST066}
\wt{\cal{L}}(\phi_i,\dot{\phi}_i)=\pi_i\,\dot{\phi}_i-{\wt{\cal{H}}}(\phi_i,\pi_i),
\ee
where $\wt{\cal{H}}(\phi_i,\pi_i)={\cal{H}}(\wt{\phi}_i,\wt{\pi}_i)$ and the latter one is the NC version of the Hamiltonian, Eq.(\ref{hamiltoniana21}), given by
\be
\label{ST067}
{\cal{H}}(\wt{\phi}_i,\wt{\pi}_i)=\frac{\wt{\pi}_i\wt{\pi}_i}{2} +\frac{(\nabla\wt\phi_i)(\nabla\wt\phi_i)}{2}+\frac{\mu^2}{2} \wt{\phi}_i\wt{\phi}_i+ \frac{\lambda}{4}(\wt{\phi}_i\wt{\phi}_i)^2.
\ee
The Hamiltonian density in Eq.(\ref{ST067}), with the help of Eq.(\ref{q-trans1}), renders to
\be
\label{hamiltonianalp41}
\wt{\cal{H}}(\phi_i,\pi_i)=\frac{\pi_i\pi_i}{2} + \frac \theta2\pi_i\varepsilon_{ij}\phi_j+\frac {(\nabla\phi_i)(\nabla\phi_i)}{2} + \wt{U},
\ee
where 
\ba
\label{potential2}
\wt{U}&=&\frac{\mu^2}{2}\phi_i\phi_i+\frac{\lambda}{4}(\phi_i\phi_i)^2+\frac18\theta^2 \phi_i\phi_i,\nonumber\\
&=&\frac{\wt\mu^2}{2}\phi_i\phi_i+\frac{\lambda}{4}(\phi_i\phi_i)^2,
\ea
with
\be
\label{newmass}
\wt\mu^2=\mu^2+\frac14\theta^2.
\ee
Observe that the original model is restored when $\theta$ is a null quantity.
 
Occasionally, energy density might be written as being the sum of kinetic and potential energy\cite{coleman},
\be
E=T+V,
\ee
where, in the Eq.(\ref{hamiltonianalp41}), $T$ is the two first term and $V$, as usual, is the term involving no canonical momenta, namely,
\be
V=\frac{(\nabla\phi_i)(\nabla\phi_i)}{2}+\wt{U}.
\ee
As a consequence, if the energy is to be bounded below, $\wt{U}$ must be also bounded below. 

The Hamilton's equation of motion $\left(\dot\phi_i=\frac{\partial \wt{\cal{H}}(\phi_i,\pi_i)}{\partial\pi_i}\right)$ is calculated and the canonical momenta are obtained as being
\be
\label{HEquation}
\pi_i= \dot\phi_i-\frac 12\theta\varepsilon_{ij}\phi_j.
\ee
Inserting Eq.(\ref{hamiltonianalp41}) and Eq.(\ref{HEquation}) into the NC first-order Lagrangian in Eq.(\ref{ST066}), we get the NC second-order Lagrangian
\be
\label{Covlagrangian}
\wt{\cal{L}}=\frac 12 (\partial_{\mu}\phi_i)(\partial^\mu\phi_i) -\frac \theta2 (n^\mu\partial_{\mu}\phi_i) \varepsilon_{ij}\phi_j-\frac{\mu^2}{2}\phi_i\phi_i-\frac{\lambda}{4}(\phi_i\phi_i)^2.
\ee
Note in Eq.(\ref{Covlagrangian}) the time-like vector $n^\mu= (1,\boldsymbol{0})$, which is a normal vector of a noncovariant set of equitemporal surfaces ($t$ = constant) where the Hamiltonian analysis is implemented. However, this noncovariance is apparent, because if we consider a larger set of space-like surfaces to develop the Hamiltonian formalism, this obstruction can be removed\footnote{This observation is well clarified by one of us in the appendix B of Ref.\cite{WN}}. From this point of view, $\theta\varepsilon_{ij}$ appear as a set of Lagrange multipliers that imposes the velocity dependent constraint $(\partial_{\mu}\phi_i)\phi_j$. As pointed out by some authors\cite{BST,BHZ,WuZee}, a Lagrangian, first-order in velocity $(\dot\phi_i)$, can always be considered as arising from a $U(1)$ background potential in configuration space. At this point, we would like to point out that the middle term of the right hand side of this NC Lagrangian, Eq.(\ref{Covlagrangian}), plays the role of damped term, \textit{vide} subsection 2.1 in Ref.\cite{MACN1}.

At this point, it is important to notice that the symmetry $(\phi_i=-\phi_i)$, previously present in the commutative global $O(2)$ scalar theory, is preserved in the NC version of global $O(2)$ scalar theory.  Further, we would like to point out that the NC transformation, given by the first equation in Eq.(\ref{q-trans1}), does not break the translational invariance in the NC second-order Lagrangian, given in Eq.(\ref{Covlagrangian}).

\section{Spontaneous symmetry breaking}
\label{sec:2}

In order to show the role played by the noncommutativity into the SSB and EBHGHK mechanism, consider the potential $\wt{U}$, given in Eq.(\ref{potential2}). In this scenario, if $\wt{\mu}^2>0$ then the vacuum is at $\phi_i=0$, the symmetry $\phi_i\rightarrow -\phi_i$ is manifest and $\wt{\mu}^2$ is the mass of the scalar modes. On the other hand, if $\wt{\mu}^2<0$, the spontaneous symmetry broken and the vacuum is at $(\phi_i\phi_i)={\frac{-\wt{\mu}^2}{\lambda}}$. In the NC framework, the usual discussion about the spontaneous symmetry broken is still valid. In order to illustrate the discussion above, we consider the following plane section, $\phi_1=0$, namely:

\begin{figure}[H]
	\centering
	\subfloat[\texttt{$\wt{\mu}^2>0$}]{
		\includegraphics[width=4.5cm]{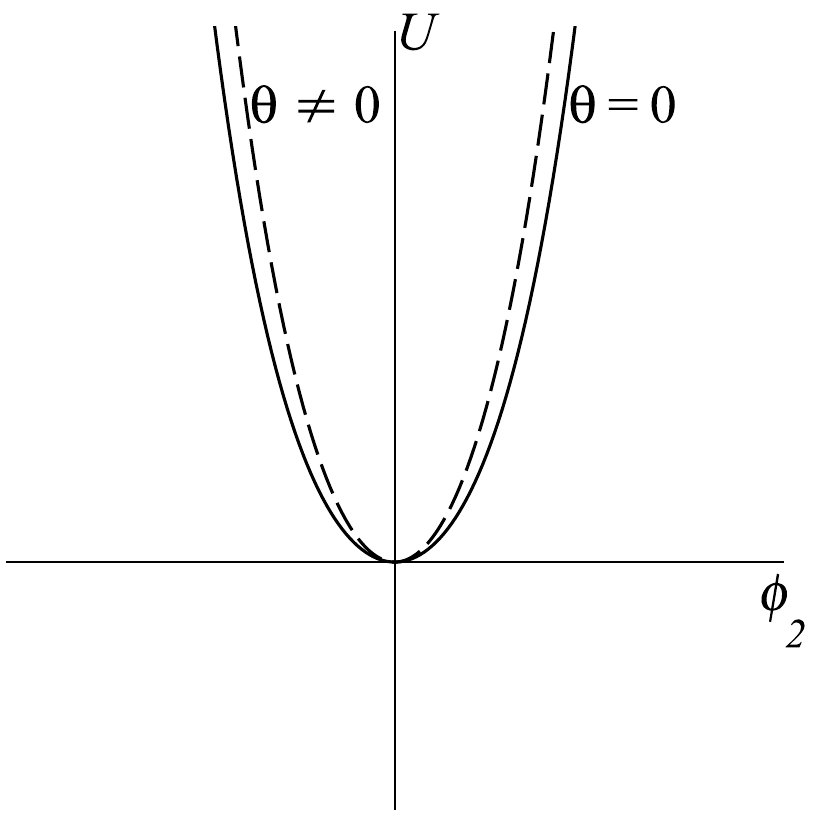}
    \label{fig:phi41}
		}
		\quad
		\subfloat[\texttt{$\wt{\mu}^2<0$}]{
		\includegraphics[width=4.5cm]{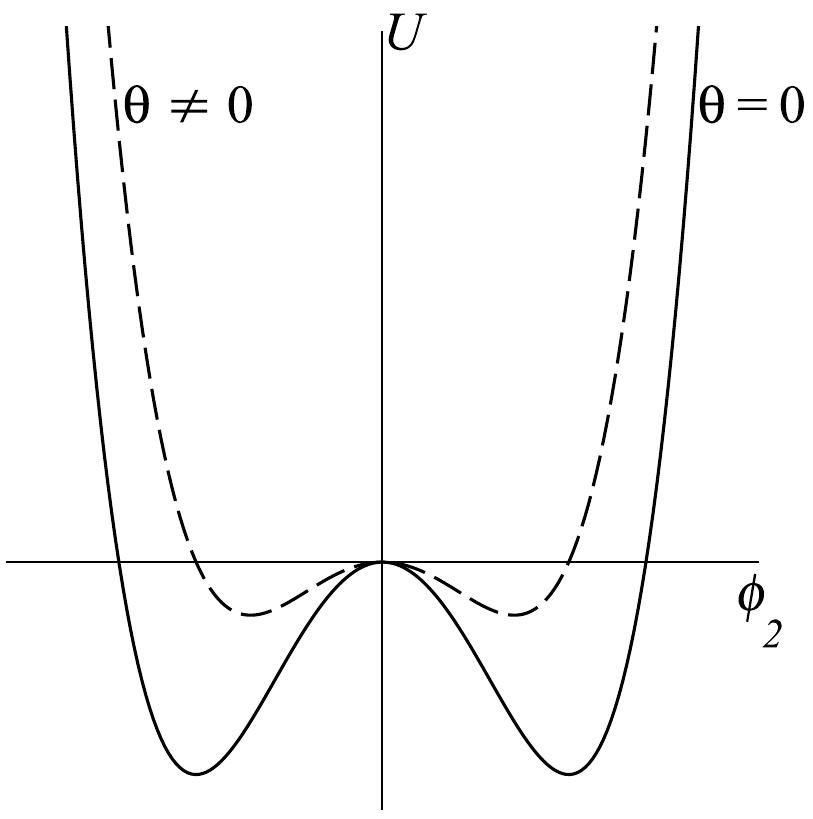}
    \label{fig:phi42}
		}
	\caption{The potential dependent on NC parameter and damping.}
	\label{fig:phi4}
\end{figure}
\noindent In Fig.(\ref{fig:phi41}), we can infer that the NC potential $(\wt{U})$ -- the dash line -- has its value increased when compared with the commutative one $({U})$ -- solid line --  due to the NC $\theta$-parameter. In Fig.(\ref{fig:phi42}) the spontaneous symmetry broken is affected by the NC $\theta$-parameter: at the bottom of the graph, the depth of the well is smaller than the one given in the commutative framework and, for any value assigned to $\phi_2$, the NC potential has its value increased when compared to commutative one. In analogy with what was done to explain the deep inelastic scattering phenomenon through DHO\cite{FT1,FT2}, where internal energy of nucleons increase after the nucleon collision so that the potential energy also increases, and together with demonstration that the NC $\theta$-parameter plays the role of a damping coefficient\cite{MACN1}, we argue that this role played by the NC $\theta$-parameter remains in the spontaneous symmetry broken context. 

In order to put our work in an appropriated way, we consider the isomorphism between $SO(2)$ and $U(1)$ group, which is implemented through the following transformation
\ba
\label{diff}
\phi_1&=&\frac1{\sqrt{2}}\,(\phi+\phi^\ast), \nonumber\\
\phi_2&=&-\frac{i}{\sqrt{2}}\,(\phi-\phi^\ast),
\ea
where $\phi$ is a complex field and $\phi^\ast$ is its conjugated one. Inserting those transformation above into Eq.(\ref{Covlagrangian}), we get
\ba
\label{Lsplitted}
\wt{\cal L}&=&(\partial_\mu\phi^\ast)(\partial^\mu\phi)-\imath\frac\theta2\eta^\mu\left[ (\partial_\mu\phi)\phi^\ast-(\partial_\mu\phi^\ast)\phi\right] \nonumber\\
&-&\mu^2\phi^\ast\phi-\lambda(\phi^\ast\phi)^2.
\ea
 
Applying the usual Legendre transformation in Eq.(\ref{Lsplitted}), the Hamiltonian is computed and its given by
\be
\wt{\cal H}=\pi^\ast\pi+(\nabla\phi^\ast)(\nabla\phi)-i\frac\theta2(\pi\phi-\pi^\ast\phi^\ast)+\wt{U},
\ee
where the potential $\wt{U}$ is 
\be
\label{sb1}
\wt{U}=\wt\mu^2\phi^\ast\phi+\lambda(\phi^\ast\phi)^2,\
\ee
with
\be
\label{sb2}
\wt\mu^2=\mu^2+\frac{\theta^2}4.
\ee
Thus the VEV for $\wt\mu^2>0~$ is 
\be
\langle\phi\rangle_0=0,
\ee
the symmetry is preserved and the mass of the complex scalar mode is given by Eq.(\ref{sb2}). 

On the other hand, the VEV for $~\wt\mu^2<0$ is given by
\be
\label{vacuum}
\langle\phi\rangle_0=\frac{v}{\sqrt{2}}
\ee
where
\be
\label{vev}
v=\sqrt{\frac{-\wt\mu^2}{\lambda}}.
\ee
It is important to notice that the VEV is already NC $\theta$-dependent, \textit{i.e}, we get a vacuum degeneracy NC $\theta$-dependent.

The field redefinition, around $\langle\phi\rangle_0$, for SSB is given by
\be
\label{transf1}
\phi=e^{i\,\zeta/v}\left(\frac{v+H}{\sqrt{2}}\right).
\ee
Therefore, the Lagrangian, given in  Eq.(\ref{Lsplitted}), renders to
\ba
\label{L1000}
\wt{\cal L}
&=&\frac12\left[(\partial_{\mu}H)(\partial^{\mu}H)+(\partial_{\mu}\zeta)(\partial^\mu\zeta)\left(1+\frac{H}{v}\right)^2\right]\nonumber\\
&+&\frac\theta2\frac{\dot\zeta}{v}(v+H)^2 - \frac 14\lambda H^4 
- \lambda vH^3-\frac 12 \left(\mu^2+3\lambda v^2\right) H^2 \nonumber\\
&-&\left(\mu^2v+\lambda v^3\right)H -\frac 12 \mu^2v^2 -\frac 14 \lambda v^4.
\ea
The spectra has one massless Goldstone boson $(\zeta)$ and one massive Higgs boson field $(H)$. Further, note that the noncommutative parameter $\theta$ leaving its trace on the VEV and the Lagrangian has a new coupling term between $\zeta$ and $H$, which is a NC $\theta$-parametrized damping term\cite{MACN1}. Further, we can read the squared mass of the Higgs boson as being
\ba
\label{mhiggs}
m_H^2&=&\mu^2+3\lambda v^2,\nonumber\\
&=& M^2-\frac{3}{4}\theta^2,
\ea
where $\,M\equiv+\sqrt{-2\mu^2}\,$ is a positive mass parameter and $\theta^2\geq0$. The Higgs mass present a degeneracy and it is NC $\theta$-dependent. Note that for $\theta=0$, the usual Lagrangian(without Higgs coupling), given in the literature\cite{coleman,IZ,TL,kaku,ryder}, is restored as well as the Higgs mass, Eq.(\ref{mhiggs}).
Considering the latest CMS experimental result for the Higgs boson mass, $(125,38 \pm 0,14)$~GeV\cite{cms-latest} for $\theta=0$; this could be also assumed the latest  ATLAS experiment result\cite{atlas-higgs}, namely, $(124,79 \pm 0,24)$~GeV. At this point, it is possible to interpret the origin of Higgs boson mass degeneracy, Eq.(\ref{mhiggs}), as being a contribution comes from the interaction with particles, even with a second-generation particles\cite{CMS2a,CMS2b,CMS2c}. In the present case, between $\zeta$ and $H$, so

\be
\label{mass-degenerate}
m_H=\sqrt{(125,35~\text{GeV})^2-\frac34\theta^2}.
\ee
The Higgs boson is the quantum manifestation of the Higgs field and, now, a NC degeneracy was introduced and this might be due to the Higgs coupling. Further, this also might be interpreted from the decay point of view. By measuring the rate at which the Higgs boson decays into different particles, the strength of their interaction with the Higgs field can be inferred: if it is higher the rate of decay into a given particle, then it is stronger its interaction with the field and \textit{vice-versa}, which is now parametrized by the $\theta$-parameter; note that this phenomenon is analogous with damping phenomenon discussed in the introduction \ref{sec:intro}. Up until now, it was measured the Higgs boson decays into different types of bosons(W\cite{atlasW,cms5} and Z\cite{atlas3,atlas4}), heavier fermions(tau leptons)\cite{cms-tau} by the ATLAS and CMS experiments\cite{atlas-cms}, heaviest quarks(top\cite{higgs-top} and bottom\cite{higgs-botton,atlasB})and muons\cite{CMS2b,atlas-muon,CMS2c,CMS2a} and these results follow the decay rate, or in an equivalent way, the damping phenomenon feature.

\section{Englert-Brout-Higgs-Guralnik-Hagen-Kibble mechanism}
\label{sec:3}
At this section, it is demonstrated how the NC $\theta$-parameter affects the mass gained by the gauge field. The Lagrangian invariant under global gauge transformation is given by
\ba
\label{Lag}
\wt{\cal L}&=&(\partial_\mu\phi^\ast)(\partial^\mu\phi)-i\frac\theta2\eta^\mu\left[ (\partial_\mu\phi)\phi^\ast-(\partial_\mu\phi^\ast)\phi\right]\nonumber\\
&-&\mu^2\phi^\ast\phi-\lambda(\phi^\ast\phi)^2.
\ea
The VEV for $~\wt\mu^2<0$ was calculated in Eq.(\ref{vacuum}) and the field redefinition for SSB, taking as starting point $\langle\phi\rangle_0$, is given by
\be
\label{fieldredef}
\phi=e^{i\,\zeta/v}\left(\frac{v+H}{\sqrt{2}}\right).
\ee
We may express the field derivative as
\be
\label{der}
\partial_{\mu}\phi=e^{i\,\zeta/v}\left[\partial_{\mu}+i\,\frac1{v}(\partial_{\mu}\zeta)\right]\left(\frac{v+H}{\sqrt{2}}\right).
\ee
The change to a new Lagrangian which is invariant under $U(1)$ local gauge transformations is implemented by
\ba
\label{Lag_gauge}
\wt{\cal L}&=&-\frac14F_{\mu\nu}F^{\mu\nu}+({D}_\mu\phi)^{\ast}({D}_\mu\phi)
-i\frac\theta2\left[(D_0\phi)\phi^\ast\right.\nonumber\\
&-&\left.(D_0\phi)^\ast\phi\right]
-\mu^2\phi^\ast\phi-\lambda(\phi^\ast\phi)^2,
\ea
where $F_{\mu\nu}$ is the field strength and $D_\mu$ is the Abelian covariant derivative defined as
\ba
F_{\mu\nu}&=&\partial_{\mu} A_\nu-\partial_\nu A_\mu, \\
{D}_\mu\phi&=&(\partial_{\mu}+ieA_\mu)\phi.  \label{covder}
\ea 
We can readily verify with the help of the Eq.(\ref{fieldredef}), Eq.(\ref{der}) and Eq.(\ref{covder}), that the Abelian covariant derivative can be rewritten as
\be
\label{covder2}
{D}_\mu\phi=e^{i\,\zeta/v}\left[\partial_{\mu}+ie\left(A_\mu+\frac1{ev}\,\partial_{\mu}\zeta\right)\right]\left(\frac{v+H}{\sqrt{2}}\right).
\ee

In order to demonstrate that the Lagrangian, given in Eq.(\ref{Lag_gauge}), is invariant under the following $U(1)$ local gauge transformation:
\be
\label{gauged}
\phi{~\rightarrow~}\phi^\prime=e^{-i\zeta/v}\phi  ~~~\text{and}~~~A_\mu{~\rightarrow~}A^\prime_\mu=A_\mu+\frac1{ev}\partial_{\mu}\zeta,
\ee
consider the gauged covariant derivative and field
\ba
{D}^\prime_\mu&=&\partial_{\mu}+ieA^\prime_\mu,  \\
\phi^\prime&=&\frac{v+H}{\sqrt{2}}.
\ea
With these latter expressions, the Eq.(\ref{covder2}) can be readily rewritten as
\be
{D}_\mu\phi=e^{i\,\zeta/v}{D}^\prime_\mu\phi^\prime.
\ee
That is, the covariant derivative of the field undergoes exactly the same transformation of the field, shown in the first Eq.(\ref{gauged}), which guarantees the invariance of the Lagrangian given in Eq.(\ref{Lag_gauge}) under local gauge transformation. Since the invariance under local gauge was guaranteed, we can, alternatively, to express the Lagrangian in terms of the primed fields or non-primed fields. Choosing express it in terms of the primed fields:
\ba
\phi^\prime&=&\frac{v+H}{\sqrt{2}}, \\
{D}^\prime_\mu\phi^\prime&=&\frac1{\sqrt{2}}\left[\partial_{\mu}H+ieA^\prime_\mu(v+H)\right],
\ea
the Lagrangian, given in Eq.(\ref{Lag_gauge}), renders to
\ba
\label{Lag_gauge_2}
\wt{\cal L}
&=&-\frac14F^\prime_{\mu\nu}F^{\prime\mu\nu}+\frac12e^2(v+H)^2A^\prime_\mu{A^\prime}^\mu+\frac\theta2\,eA^\prime_0\,(v+H)^2\nonumber\\&+&\frac12(\partial_{\mu}H)(\partial^{\mu}H)-\frac 14\lambda H^4 - \lambda vH^3-\frac12\left(\mu^2+3\lambda v^2\right) H^2 \nonumber\\&-&\left(\mu^2v+\lambda v^3\right)H -\frac 12 \mu^2v^2 -\frac 14 \lambda v^4.
\ea
The degree-of-freedom previously associated with the massless Goldstone boson was transferred to the longitudinal sector of the gauge field making the latter massive. Further, note that the noncommutative $\theta$-parameter leaving its trace on the VEV and the Lagrangian has a new coupling term$(\theta\texttt{-dependent})$ between the scalar potential $A^\prime_0$ and Higgs field. We also can read the squared mass of the gauge field  as being
\ba
\label{mAtheta}
m_{A}&=&ve\nonumber\\
&=& \sqrt{\left({M^2}-\frac{\theta^2}2\right)\frac{e^2}{2\lambda}}, \nonumber\\
&=& \sqrt{\left({(125,35~\text{GeV})^2-\frac54\theta^2}\right)\frac{e^2}{2\lambda}}, \label{m_gauge} 
\ea
where Eq.(\ref{sb2}), Eq.(\ref{vev}) and Eq.(\ref{mass-degenerate}) were used. Note that for $\theta=0$, the usual Lagrangian given in the literature is restored.

\section{Conclusion}
\label{sec:4}

We would like to point out that a NC $O(2)$ scalar field theory was obtained through a non-Moyal $\ast$-product and it works as being a damped field theory and does not present a no-go problem. Further, it was revealed an astonishing feature about the role played by noncommutativity in the SSB and EBHGHK mechanisms: first, there is a new Higgs coupling term, NC $\theta$-dependent, in the model; second, the vacuum state presents a degeneracy due to the NC $\theta$-parameter: the VEV is now NC $\theta$-parameter depended or a damping term is presented \textit{ab initio}, \textit{i.e.}, the universe should present a NC nature. Remember that the Higgs field is presented throughout all space, then the vacuum of outer space is not empty but contain this field. Consequently, Higgs couplings are already presented and, since these couplings play the role of a damping phenomenon associated to a NC $\theta$-parameter, the universe should present a NC nature. Due to this, it is possible to argue that the particle set coupling to Higgs has already a NC $\theta$-parameter dynamical mass before SSB; third, and after SSB, the vacuum degeneracy(due to damping) drives the degeneracy to the Higgs boson mass, \textit{i.e.}, the Higgs mass is now a NC $\theta$-parameter depended. This, maybe, might shed some light on the nature of how Higgs field is coupling to the other particles before SSB; maybe, the couplings of Higgs field are due to a set of mediating particles, which may explain the NC behavior of universe. The Higgs boson mass is parametrized by the NC $\theta$-parameter and, therefore, we get a NC $\theta$-parameter dependent mass-degenerate Higgs boson near 125~GeV after SSB, namely, $m_H=\sqrt{(125,35~\text{GeV})^2-\frac34\theta^2}$ with $\theta$ as being a parameter that also represent the new Higgs coupling in the model. This opens up the possibility that the theoretical result can be settle to the phenomenological one; fourth, a new coupling term between Higgs field and gauge vector field appears, whose coupling parameter is $\theta$-dependent. Moreover, the gauge vector field gains a mass parametrized by the NC $\theta$-parameter,  {\textit{vide}} Eq.(\ref{mAtheta}), which can be fitted to experimental results by simply tuning the NC $\theta$-parameter. 

\section*{Acknowledgments}
 
C. Neves and M. A. De Andrade would like to thank Brazilian Research Agencies(CNPq and FAPERJ) for partial financial support. Further, we would like to thank Bruno Fernando Inchausp Teixeira for a careful reading of the manuscript.

\end{document}